\documentclass[showpacs,twocolumn,aps]{revtex4}
\usepackage{graphicx, epsfig}

\newcommand{\be}{\begin{equation}}
\newcommand{\ee}{\end{equation}}
\newcommand{\bea}{\begin{eqnarray}}
\newcommand{\eea}{\end{eqnarray}}

\begin{document}

\title{Can we predict $\Lambda$ for the Non-SUSY sector of the Landscape ?}

\author{Laura Mersini-Houghton}

\affiliation{Department of Physics and Astronomy, UNC-Chapel Hill, NC 27599-3255, USA}

\date{\today}

\begin{abstract} 
We propose a new selection criteria for predicting the most probable wavefunction of the universe that propagates on the string landscape background, by studying its dynamics from a quantum cosmology view. Previously we applied this proposal to the $SUSY$ sector of the landscape. In this work the dynamic selection criterion is applied to the investigation of the non-$SUSY$ sector.In the absence of detailed information about its structure, it is assumed that this sector has a stochastic distribution of vacua energies.The calculation of a distribution probability for the cosmological constants $\Lambda_{eff}$, obtained from the density of states $\rho$, indicates that the most probable wavefunction is peaked around universes with zero $\Lambda_{eff}$. In contrast to the {\it extended wavefunction} solutions found for the $SUSY$ sector with $N$-vacua and peaked around $\Lambda_{eff}\simeq \frac{1}{N^2}$, wavefunctions residing on the non-$SUSY$ sector exhibit {\it Anderson localization}.Although minisuperspace is a limited approach it presently provides a dynamical quantum selection rule for the most probable vacua solution from the landscape.
\end{abstract}


\maketitle

Recently we proposed to use quantum cosmology on the background of the landscape of string theory in order to find the most probable wavefunction of the universe\cite{Kobakhidze:2004gm}. This proposal provides a new selection criterion for predicting the most probable wavefunction of the universe based on the dynamics of the superspace. In \cite{Kobakhidze:2004gm} we focused our attention on the minisuperspace of the supersymmetric (SUSY) vacua sector of the landscape. The SUSY and non-SUSY sectors of the landscape were assumed disconnected.

Here we investigate solutions for the most probable wavefunction of the universe by extending our previous analysis to the non-SUSY sector of the landscape. The minisuperspace approach is a reasonable approximation for both SUSY and non-SUSY sectors because, as shown below, the previous assumption that the two sectors are disconnected is fully justified. The purpose of this work is to examine the question: Can we make a prediction for the most probable and stable solution for the wavefunction of the universe propagating on the background of the non-SUSY sector of the landscape?

\paragraph*{Preliminaries and Setup:} Consider the array of landscape vacua solutions in the non-SUSY sector to be a crystal.
The array of vacua are parametrized by the potential $V(\phi)$ of the moduli field $\phi$ which collectively characterizes all the relevant degrees of freedom of string fluxes and fields $\phi= [\Phi_n ]$. The vacua are located at $\phi_i$, $i=1,...,N$. The energies of the non-$SUSY$ vacua, $\lambda(\phi_i)=\lambda_i$, can take any value in the range $\lambda_i \in \left[0, \pm W \right]$ where $W =O(M_p^{4})$ is an energy  scale related to the Planck or string constant scale. The {\it crucial assumption} is that the vacua energies on the non-$SUSY$ sector are stochastic, namely $\lambda_i$'s are drawn randomly in the interval $\left[0, \pm W \right]$. The spacing between different vacua could also be a random 'sprinkling' of sites. In such a setup, the non-$SUSY$ sector of vacua is a randomly disordered lattice of configuration (super-)space of moduli $\phi$, which we name the 'superlattice'. The goal is to use quantum cosmology on the landscape background for dynamically selecting the most probable wavefunction of the universe propagating on the 'disordered superlattice'.

\paragraph*{Wavefunction of the Universe on the Stochastic Minisuperspace of the non-$SUSY$ Sector:} 
Following the proposal of\cite{Kobakhidze:2004gm} let us define the non-$SUSY$ minisuperspace to be restricted to the 'disordered superlattice' of the stochastic non-$SUSY$ sector $V(\phi)$ of vacua $\phi$ and to homogenous flat $3-$geometries
\begin{equation}
ds^2= \left[-{\cal N}dt^2+a^2(t)d\bf{x}^2\right],
\label{1}
\end{equation}  
with ${\cal N}$ being the lapse function. The non-$SUSY$ minisuperspace is parametrized by the variables, $[a,\phi]$. The Lagrangian 
for the system with variables $[a,\phi]$ receiving contributions from both: gravity ($L_g$) and, moduli 'superlattice' ($L_{\phi}$), is $L=L_g+L_{\phi}$ where
\begin{eqnarray}
L_g=-\frac{3M_p^2}{8\pi}\frac{a\dot{a}^2}{{\cal N}} \nonumber \\
L_{\phi}=\frac{a^3{\cal N}}{2}\left(-\frac{\dot{\phi}^2}{{\cal N}}-V(\phi)\right)\nonumber\\
\end{eqnarray} 

$\Psi[a,\phi]$ denotes the wavefunctional of the universe propagating on the minisuperspace background $[a,\phi]$.
 The Hamiltonian constraint on the wavefunctional is obtained by varying the combined action with respect to the lapse function $\cal N$.In the usual manner\cite{wdw,Hartle:1983ai,Vachaspati:1988as}, promoting $p_a$, $p_{\phi}$ into operators,gives the Wheeler-De Witt (WDW)\cite{qcreview,wdw} 
equation
\begin{eqnarray}
{\hat {\cal H}}\Psi(a,\phi) = 0 ~{\rm with} \nonumber \\
\hat{{\cal H}}=\frac{1}{2e^{3\alpha}}\left[\frac{4\pi}{3M_p^2}
\frac{\partial^2}{\partial\alpha^2}-
\frac{\partial^2}{\partial\phi^2}+e^{6\alpha}V(\phi)\right] \nonumber \\
\label{2}
\end{eqnarray}
where $a$ is replaced by $a=e^{\alpha}$. (A treatment of the subtleties related to the normal ordering of operators
and cross-terms in the Wheeler-DeWitt equation, Eqn. (3), can be found in Ref.[28,29,30])

Rescaling $\phi$ to $x=e^{3\alpha}\phi$, (and the other relevant quantities in the potential), in order to separate 
variables in (\ref{2}), $\Psi[\alpha , \phi]$ can be decomposed in modes
\begin{equation}
\Psi({\alpha, x})=\Sigma _j \psi_j(x)F_j(\alpha).
\label{3}
\end{equation}
Replacing (\ref{3}) into (\ref{2}) and using 

\begin{eqnarray}
\hat{\cal{H}}(x)\psi_{j}(x) = \hat \epsilon_j \psi_{j}(x) ~{\rm where} \nonumber  \\
\hat{\cal{H}}(x)=\frac{3M_p^2}{4\pi} \left[
\frac{\partial^2}{\partial x^2}- V(x)\right] \
\nonumber \\
\label{41}
\end{eqnarray}

results in 
\begin{equation}
-\frac{\partial^2}{\partial \alpha ^2}\psi_j(x)F_j(\alpha) = \hat{\epsilon_j}\psi_jF_j.
\label{4}
\end{equation}
where 'hat' denotes the rescaled $\hat{\epsilon_{j}} = e^{6\alpha} \epsilon_j$ and from hereon$\frac{3M_p^{2}}{4\pi}$ is absorbed into $\epsilon_j$ given in fundamental units.
The $\alpha$ equation of motion is obtained by varying the action, $S$ resulting from Eqn. (\ref{2}) with respect to $\alpha$
\begin{equation}
\ddot{\alpha}+\frac{3}{2}\left[\dot{\alpha}^2+(\dot{x}^2-V(x))e^{-6\alpha}\right]=0 .
\label{alpha}
\end{equation}   
It is easy to check for consistency that the $\alpha$ equation of motion is the Friedman equation for the expansion

 The 'eigenvalues' $\hat{\epsilon_j}$ are obtained by solving the 'Schrodinger' type equation, Eqn.(~\ref{41}), for 
the field $\psi_{j}(x)$ propagating on the disordered superlattice $V(x)$ with $N$ lattice sites 
${x_i}$ and stochastic vacua energies $\lambda_i$. The hamiltonian $H(x) = H_{0} + H_{I}$ contains two pieces:
the diagonal part $H_{0} = \frac{\partial^2}{\partial x^2} -V_{0}$  where $V_{0}(x_i)=\lambda_i$ and the short range interaction between neighbor vacua [$x_i,x_j$], (the nondiagonal terms), $H_I = V_{I}$.Therefore the potential $V(x)$ includes two terms $V(x)=V_0+V_{I}(x)$.  The short range interaction allows spreading of the wavefunction. Let us assume the nearest neighbor approximation for the short range interaction, for example think of tuneling to the nearest neighbors for simplicity, and introduce a (dimensionless) $\delta$-correlated white noise for the interaction term, i.e $< V_I >=0$ and $< V_{I}(x_i) V_{I}(x_j)> =\Gamma \delta(x_i - x_j)$ where $<...>$ denotes ensemble averaging with respect to [$x_j$]. (Note that we do not assume $<V_0>=\bar\Lambda$ to be centered around zero). Due to stochasticity the potential and eigenvalues can not be written in an exact form. In most cases, the details of $V(x)= V_0 +V_I$ do not matter since all the results are obtained only in a probabilistic manner.This is a complicated $N-$body problem because of the multiscattering of $\psi_j(x)$ among many sites of the 'superlattice' but relevant quantities are calculated from probability distributions which can be found exactly.  

There is a vast amount of literature 
on the propagation of wavefunctions on disordered lattices and the treatment of solutions for types of Eqn.(\ref{41}) can be found in many papers,\cite{anderson,review1,review2,review3,efemetov}. P. Anderson was awarded the Nobel 
prize for his investigation into this issue. Independent of the specifics of the model, the conclusion is that the wavefunctions with energies below a certain scale set by the disorder strength do not propagate 
on random lattices with short range interactions. Instead, the wavefunction soon gets localized around a lattice site $x_j$ with 
vacuum energy $\lambda_j$, a phenomenon known as Anderson localization\cite{anderson}.Localization is purely a quantum mechanical effect and it occurs because of the destructive interference of the phases of the wavefunction from multiple scattering among sites. As shown in [3], this phenomenon always occurs when: disorder is present and, the interaction term $V_I$ is sufficiently short-range, for all levels with energy below the disorder strength set by $\Gamma$. The $1D$ and $2D$ disordered systems are critical in the sense that localization occurs for all energy levels, independent of the disorder strength $\Gamma$. Many examples of this phenomenon have been succesfully applied to lattice QCD\cite{qcd}, observed experimentally in condensed matter systems\cite{observed} and later derived from the (RMT) Wigner-Dyson theory of random matrices\cite{review1,review2,review3,efemetov,hole1,hole2}, with the $N X N$ matrices obtained over many realizations of the random potential\footnote{ The 1 and 2-dimensional lattices are special cases as localization always occurs even for weak disorder}. Unfortunately there is a no one-to-one correspondence between the localization site and the energy, since vacuum energies are stochastic variables with values assigned randomly from the interval $[0 ,\pm W]$, and $W=O(M_p^4)$ for our case. However, conclusions can be drawn about the energy of the localized wavefunctions from their distribution probability shown below. 

Let us consider the distribution function of the vacua energies from the interval $[0,\pm W]$ by 
$P(\lambda)$. $P(\lambda)$  and therefore $P(\hat{H}(x))$ can be a Gaussian distribution $P(\lambda)=\frac{1}{\sqrt{N\Gamma}}
e^{-\frac{(\lambda-\bar{\Lambda})^2}{N\Gamma}}$, with ensemble averaged mean values $<1|\Lambda_i|N>
=\bar{\Lambda}$ and width $\Gamma =<1|\Gamma_{i}|N>$, or; when disorder is large $\Gamma =O(W)$  
by a flat distribution $P(\lambda)=\frac{1}{2W}$. 
Corrections to the unperturbed energy $\lambda_j$ for the wavefunction localized around $x_j$, along with the evaluated Green's function $<j|G|j> =<j|( \epsilon - \hat{H}_{x} )^{-1}|j>$ can be estimated by the usual perturbation theory \cite{anderson} for weak disorder and by Wigner-Dyson RMT methods for the case when disorder is large and perturbation theory breaks down. The latter is a more elegant and transparent method of calculation since averaging $<...>$ is done by integrating with respect to $\hat{H}(x)$ with probability weight $P[\hat{H}(x)]\simeq e^{ -\frac{{\hat{H}(x)}^{2}}{N\Gamma} }$ \cite{efemetov} rather than ensemble averaging over vacua $[x_j]$. The case of lattices with stochastic energy distributions can be found in many reviews\cite{efemetov,review1,review2,review3,anderson} thus below we outline only the main steps of the calculation relevant to our purposes. Let us denote the shift of the unperturbed energy at 'site' $x_j$ by $\Sigma_j = \delta_j + i\gamma$. Perturbed energy $\epsilon_j$ and Green's function are
  
\begin{eqnarray}
\epsilon_j\approx\lambda_j-\delta_{j}
- i\gamma \nonumber \\
G^-1_{jj}=(\epsilon - \epsilon_j) \nonumber \\
\label{energy}
\end{eqnarray}      
with $\delta_j$ the usual second-order correction that for simplicity we may take to vanish. With perturbation theory methods 
$i\gamma_j=\frac{i\Gamma_j}{\epsilon_j}=i\Sigma_{k\neq j}\frac{|V_{I_{jk}}|^2}{\epsilon_j}$, where $V_{I_{jk}}$ denotes
interaction between sites $x_j$ and $x_k$. The same results are obtained by calculating $\Sigma_j$ with the Wigner-Dyson RMT formalism,(for techincal reasons up to self-energy corrections in most cases), $\gamma_j \simeq \frac{2\pi\Gamma}{\epsilon_j}$.  Assuming that, independently of the 'site' $x_j$ (white noise), the second correlation 
probability momentum falls rapidly fast away from $x_j$ then the most probable and averaged mean value for $\Gamma_j$ are roughly the same. Thus we replaced it with its mean value $\Gamma=<1|\Gamma_j|N>$, in the expression $<V(x_j)V(x_k)>\approx \Gamma\delta(x_j-x_k)$.). 

The eigenfunction $\psi_j(x)$ localized around $x_j$ is given by 
\begin{equation}
|\psi_j(x)|^2\sim \frac{1}{l_j}e^{-\frac{x-x_j}{l_j}}
\label{loc}
\end{equation}
Due to Anderson localization $\psi_{j}(x)$ can not propagate from the non-$SUSY$ sector to other sectors of the landscape. (Therefore the assumption that the two sectors are disconnected seems fully justified). Localization is a demonstration of purely quantum mechanical effects.Classically the amplitudes of scattering/tunneling through many vacua would be multiplied over all the 'sites' $1..N$ which means that the final transition amplitude for propagation through a periodic lattice would be zero. Clearly this is not the case, (otherwise there would be no metals in nature), because quantum mechanically phases of the wavefunction add up coherently and the constructive interference of phases after the multiple scattering over many sites results in a large final transition amplitude. Thus the propagation through the periodic lattice.Introducing disorder on the periodic potential of the lattice destroys the constructive interference among the phases, which results in the phenomenon of localization.For this reason, Anderson localization is a display of quantum mechanical multiscattering effects\footnote{This small regression in the subtleties of quantum effects was made to clarify that the approach we have proposed of a quantum mechanical system $\Psi[x]$ propagating on {\it configuration space} $V(x)$ should not be confused with a quantum field calculation of tunneling of $\psi$ with potential $V(\psi$ in real {\it space-time} which indeed is suppressed by the volume of space and thus insignificantly small}.   
The averaged localization length of the system is obtained from the exponential decay of the retarded Green's function and given by the 
ensemble average of the norm of the retarded Green's function $G_R^{-1}$, ($\gamma =<\gamma_j>$), by $l=<l_j>$, $\frac{l}{L}=\frac{1}{\pi}<1|ln||G^{-1}(x_i,x_j)||N> \simeq (\frac{1}{\gamma})\simeq (\frac{2W}{\Gamma})$ where localization lengths $l_j$ are related to $\gamma_j$
by $l_j\sim \frac{1}{\sqrt{\gamma_j}}$ and $L$ is the size of the landscape sector, $L\simeq Nl_p$.

Replacing the solution Eqn.\ref{41},and Eqn.\ref{energy}, (where for simplicity $\delta_j$ is taken zero), back to WDW Eqn.(\ref{4}), we get $F_j(\alpha)\sim e^{\pm
\sqrt{\hat{\epsilon_j}}i\alpha}$.The wavefunction of the universe solution is localized in the moduli 'superlattice' around some vacua $x_j$ but it has an oscillatory behaviour with respect to $\alpha$.

\begin{equation}
\Psi_j(x,\alpha)\simeq \frac{1}{ \hat{\epsilon_j}^{1/4} \sqrt{l_j}}e^{\pm i\sqrt{\hat{\epsilon_j}}
\alpha-\frac{(x-x_j)}{2 l_j}}
\label{5}
\end{equation}  

As in \cite{Kobakhidze:2004gm} we obtain the solution for $F_j(\alpha)$ in Eqn.(\ref{4})
 by using the Vilenkin boundary condition of choosing the outgoing modes at future infinity \cite{Vilenkin:1994rn}. The time parameter is obtained from Eqn.\cite{alpha} 
\begin{equation}
\sqrt{\hat{\epsilon_j}}\alpha = {H_j}t
\label{hubble}
\end{equation}
where ${H_j}$ is the expansion rate experienced by the local observers bound to the universe $\Psi_j$.
The oscillatory behaviour needed for interpreting $\alpha$ as the time parameter is exhibited only for positive energies $\hat{\epsilon_j}$.The choice of outgoing modes boundary condition  thus picks only wavefunctions localized around sites 
with positive energy $\hat{\epsilon_j} \geq 0$ for 'giving birth' to a universe. (AdS-)solutions with $\hat{\epsilon_j}$ 
in the interval $[-W,0]$ are decaying solutions and thus physically irrelevant for this choice of boundary conditions as they can not
give rise to a universe. That is, gravity ($L_g$) in Eqn.(\ref{2}) breaks the 'particle-hole' symmetry $\epsilon \to -\epsilon$ on the 'superlattice', namely:boundary conditions and $L_g$ distinguish states of positive energy by excluding their symmetric negative energy partners\cite{hole1,hole2}\footnote{The role of different choices of boundary condition and gravity on the propagation of the wavefunction of the universe on the 'superlattice' considered here merits further study.Most current literature on localization does not consider systems coupled to gravity.}. We briefly discuss at the end why this important symmetry for the behavior of the density of states near zero energy levels is closely related to the role of the boundary conditions on the selection of the most probable wavefunction of the universe and its $\lambda$. 
\paragraph*{Density of States and the Most Probable Universe on non-$SUSY$ Sector}:
We still have to answer the question: Which solution of Eqn.(\ref{2}) is the most
probable. At first this seems a hopeless task since energies are distributed randomly in the allowed energy band. However, the question can be addressed statistically by maximizing the density of states (DoS) 
$\rho(\epsilon_i)$ . The single-particle averaged density of states can be obtained from the imaginary part of the advanced Green's function, $Im G_A  (\epsilon_j)\simeq \frac{\gamma_j}{(\epsilon-\epsilon_j)^2 +\gamma_{j}^2}$ with poles at $|\epsilon|=|\lambda_j -i\gamma_j|$ through the expression $\rho(\epsilon)=\frac{1}{\pi}<1|Im G_A|N>$  or more explicitly in RMT from $\rho(\epsilon)= \frac{1}{N}<Tr\delta(\epsilon-H(x))>_{H_{x}} = \frac{1}{N\pi} \int{D(\hat{H_{x}}) P(\hat{H_{x}})Im (G_{A})}$. For our simple 1-dimensional 'superlattice' this equation yields  

\begin{equation}
\rho(\hat{\epsilon})\approx \frac{1}{|\hat{\epsilon}| +\frac{1}{l^2}}
\label{6}
\end{equation}

The dimensionality of the 'superlattice' does not affect the main inverse power law behavior of $\rho$ on energy but it does change the power of $\epsilon, l$ in the denominator of Eqn.\ref{6}.
From Eqn.\ref{6}, the maximum is around $\hat{\epsilon_j}=\hat{\epsilon}\approx 0$, or $\lambda_j\approx  {\gamma_j}$, Fig.1. 
Note that for a highly random system, the disorder strength can be as large as ${\gamma}\sim O(\sqrt{W})$.

\begin{figure}[t]
\raggedleft
\centerline{
\epsfxsize=2.5in
\epsfbox{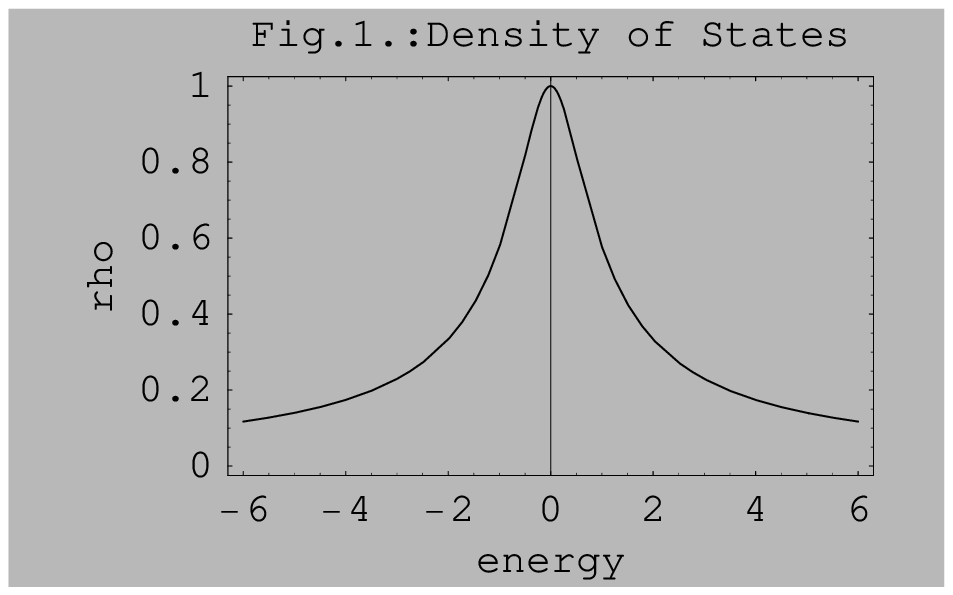}}

\label{fig:1}
\end{figure}

The {\it most probable solution for the wavefunction of the universe} $\Psi_0$ can now be found from the maximum of  
the density of states since $\rho(\epsilon)$ provides the distribution probability of states. From Eqn.\ref{6}, it can be seen that $\rho(\epsilon)$ is peaked around $\epsilon_j\approx 0$ which leads to the conclusion that, with our approach, {\it the most probable universe for the non-$SUSY$ sector}, {\it is the one with a 'physical' cosmological constant $\Lambda_{eff}=\hat{\epsilon}\simeq 0$}. Although in Fig.1 for the sake of generality, the plot of the DoS $\rho(\epsilon)$ is shown for all energies, the relevant levels here consistent with the choice of Vilenkin boundary conditions are only the positive energy levels, $\epsilon \ge 0$.

Let us define the terminology {\it physical} vs. {\it bare} here and distinguish between the 'physical' cosmological constant $\hat{\epsilon_j}=\Lambda_{eff,j}$ and the bare vacuum energy $\lambda_j = \Lambda_{bare,j}$. Our argument below is that the wavefunction of the universe experiences the self-gravitating properties of only $\Lambda_{eff}=\epsilon$ since this is the relevant energy that enters in Eqn.\ref{hubble} for the expansion of that universe. Let us think of two 'observers': a {\it 'local observer'} bound to the wavefunction of the universe $\Psi_{j}[a,\epsilon_j]$, (while oblivious about the landscape vacua energies [$\lambda_i$)], measures $\epsilon_j$ as his/her physical cosmological constant in his/her universe because it is only $\epsilon_j$ that determines the expansion rate ${H_j}$ in $F_j(\alpha)$ from $\sqrt{\hat{\epsilon_j}}\alpha = H t$; but a {\it 'superobserver'} bound to the landscape superlattice will however observe all the vacua of the landscape and 'notice' that the wavefunction $\Psi_{j}[a,\epsilon_{j}\simeq 0]$ is localized around some landscape vacuum with some large vacuum energy $\lambda_j$, (thus the name 'bare vacuum energy' for $\lambda_j$,although its perturbed energy $\epsilon_j\approx 0$). 

DoS is the most useful quantity for extracting statistical predictions about the landscape in this approach. Let us analyze briefly its main characteristics: The density of state is a maximum and does not diverge at $\epsilon =0$ due to the breaking of the 'particle-hole' symmetry by gravity. If this symmetry were preserved the divergence of $\rho(\hat{\epsilon})$ around $\hat{\epsilon}\simeq 0$ would signal either delocalization of $\psi(0)$, $l\simeq O(L=Nl_p)$ or the need to go to higher order corrections such as density-density correlations $<\delta\rho^2>$; We suspect that the density of states is a maximum around $\epsilon=0$ because of quantum effects like tuneling whereby the most localized wavefunctions are the states with the lowest energy. Probably when the wavefunction gets localized on a site $x_j$ with some large energy $\lambda_j$, because of the short range interaction or tuneling, there is a large probability that this wavepacket will still sample the nearby vacua for awhile by tunneling towards the ones with lower and lower energies until it settles to the lowest 'physical' vacua energy level\footnote{Localization length $l_j$ can be interpreted as a correlation length for the short-range interaction with the neighbor vacua since it characterizes the spread of the wavepacket around 'site' $x_j$. It is interesting to note that 
$\rho(0)\approx l^2$ seem to correspond to a vacuum 'condensate' with a large mass $m^2 \simeq O(\rho(0)^{-1})$ characterizing the lack of long-range interaction and the breaking of the 'particle-hole' $\epsilon\to -\epsilon$ symmetry. When this symmetry is preserved DoS diverges, $m^2\simeq\rho(0)^{-1}\to 0$, which signals the delocalization of the sero energy state.};DoS falls off as power-law rather than exponential towards the tail end which would give a nonnegligible probability weight to the higher energy states.

 \paragraph*{Discussion:}String cosmology is still a field in making.A lot of effort has been devoted to it based on the expectation that the field will result in predictions and provide answers for some deep challenges facing cosmology. Progress has been made in addressing puzzles like dark energy\cite{tde,bvafa}, pioneering work on string inflation\cite{Kachru:2003aw,Bousso:2000xa} or even searching for observational signatures\cite{tcmb}. However lately it was realized that progress achieved thus far soon may become a stumbling block:while rich in phenomenology, the field can provide not only one vacua solution for a universe like ours but many billions of billions of them, quite likely a whole {\it landscape} with $10^{100}$ vacua or more\cite{Banks:2003es,Douglas:2003um,Denef:2004dm,Douglas:2004zg,vafa,dvali}. The danger of losing predictability and thus becoming a nonfalsifable theory\cite{lee} places string cosmology temporarily in an uneasy position from where even appealing to anthropic selection rules may seem an attractive possibility\cite{Susskind:2003kw,Susskind:2004uv,nima}. 

The purpose of our approach, first introduced in \cite{Kobakhidze:2004gm}, was to {\it propose a new selection criteria for the landscape vacua}, based on the dynamics of the wavefunction of the universe propagating on the landscape background. We proposed to place quantum cosmology on the landscape background, thereby calculating the most probable wavefunction of the universe from solutions to the WDW equation\cite{qcreview,wdw}.Although minisuperspace is a limited aproach and issues like the role of boundary conditions and normal-ordering of operators are still under debate\cite{Hartle:1983ai,Vachaspati:1988as,Vilenkin:1994rn,Halliwell:200yc,Halliwell:2000mv,Hall:2003sm},this approach presently provides a dynamical quantum selection criterion for predicting the most probable universe with vacuum energy $\Lambda$.Initially this selection rule was applied to the $SUSY$ sector of the landscape and the most probable universes were 'standing wave' solutions extended over the whole landscape $SUSY$ sector, peaked around energies $\Lambda\simeq \frac{1}{N^2}$, with $N$ refering to the number of vacua\cite{Kobakhidze:2004gm}.

The same analysis is applied in this work to the investigation of the minisuperspace restricted to the non-$SUSY$ sector of the landscape and flat 3-geometries.In the absence of knowledge about the detailed structure of the non-$SUSY$ sector, vacua energies were considered to be a {\it stochastic} variable, namely randomly drawn from the interval $[0,\pm W]$.Wavefunction solutions found from WDW equation, Eqn.\ref{6}, exhibit the well-known phenomenon of Anderson localization characteristic of disordered systems. Localization of the wavepacket ensures that coherence of the wavefunction of the universe is maintained over large time-scales. Due to the stochastic distribution of landscape vacua energies $\lambda_i, i=1,..N$ we can not predict on which vacua of the superlattice the wavefunction $\Psi(x,\alpha)$ will be 
localized or the exact vacuum energy it will find there. However, we can make statistical 
predictions for the distribution probability of the physical cosmological constants $\Lambda_{eff,j} =\hat{\epsilon_j}$ and the most probable wavefunction of the universe $\Psi_{j}[a,\hat{\epsilon_j}]$. This information is extracted from the density of states $\rho(\hat{\epsilon_j})$. In the non-SUSY sector of the landscape our findings indicate that the most probable wavefunction of the universe solution selects 
states of zero energy and consequently of zero 'physical' cosmological constant, (although the superlatice 'bare' vacua energies $\lambda_i$ where the most probable wavefunction localizes, may be as large as the disorder strength $W$, $\lambda_j\approx O(\gamma)$). 
Statistically, the most probable 'physical' cosmological constant 'as seen by the $\Psi_j$-bound local observer' through measurement of his/her expansion rate $\Lambda_{eff,j}={H_j}=\sqrt{\hat{\epsilon_j}}$, is the perturbed energy 
$|\epsilon_j|\approx |\lambda_j\pm \sqrt{\Gamma}|$ peaked around $|\epsilon_j|\simeq 0$. 
This behavior of DoS may result from the following: as the wavefunction gets localized 
at some vacua $x_j$ with vacua energy $\lambda_j$ it can still sample nearby vacua by quantum mechanical tunneling to neighbor sites $x_k$ that have lower vacuum energies $\lambda_k$. This interaction contained in $V_I$ results in energy corrections from disorder $\gamma_j$. After 'moving around' to sample the neighbor vacua the localized wavepacket can finally settle to the site with the lowest energy. The probability of the reverse path is much smaller and thus does not dominate the DoS $\rho(\hat{\epsilon)}$.We suspect it is for this reason that the low energy states dominate the expression for $\rho(\epsilon)$. However it should be noted that the width of distribution in the plot for the DoS, in Fig.1,  is of order the disorder $\Gamma$. Therefore, depending on the strength of disorder, the probability of solutions for universes with nonzero cosmological constants may still be significant. The breaking of the 'particle-hole' $\epsilon\to -\epsilon$ symmetry by gravity ensure that DoS does not diverge at $\epsilon\to 0$ which guarantees the existence of localized solutions at zero energies\cite{hole1,hole2}.It is interesting to try to understand the relation between the existence of a localized solution around zero energies with time-reversability symmetry determined by the boundary conditions. The Vilenkin boundary condition we employed here, clearly has a preferred direction of only outgoing modes at time infinity. A different choice of boundary conditions that would mix incoming and outgoing modes thereby preserving time-reversability, would likely result in a divergent DoS near zero energies,which can be interpreted as the lack of solutions with zero physical cosmological constant from the probability view of DoS.The relation between DoS and time-reversability symmetry deserves careful examination and will be reported elsewhere since it is beyond the scope of this paper. As an aside,from the expression for DoS, Eqn.\ref{6}, it should be noted that states with $0\ge \epsilon_j \leq  \gamma$ still have a non-negligible probability since the occupation of states at the tail end of DoS is only power-law suppressed.
   
DoS plays the role of an order-parameter for the disordered 'superlattice'. Its peculiar behaviour near zero and critical energies can signal phase transitions and critical phenomena. Therefore it is a very important quantity for exploring the rich phenomenology of the landscape. Many issues remain to be addressed yet. The limited degrees of freedom for the minisuperspace we considered in this work do not allow addressing problems like the coupling of the standard model to landscape moduli and its backreaction effect on the expansion rate $H_i$, the DoS and the most probable wavefunction, or the later evolution of a universe born out of the wavefunction localized around some landscape vacua. Future directions for the application of our proposal have to involve an extension of the minisuperspace to more degrees of freedom needed for addressing more realistic scenarios, for example an extension to non-flat 3 geometries or to universes with the standard model content in them. The latter, currently under study, can be used to relate Higgs to $\lambda$ hierarchies through universal scaling and critical phenomena. We report our results about the latter in a separate publication. Despite the current limitations of quantum cosmology, we believe it is important and a step forward to provide an alternative proposal to the anthropic principle that dynamically selects the landscape vacuum in which our universe resides, as shown in this work.

{\it Acknowledgment:} I am grateful to D.Khveshchenko and L.McNeil for valuable discussions and references on condensed matter analogs. I would also like to thank A.Kobakhidze for useful discussions.

\end{document}